\documentclass[
 reprint,
superscriptaddress,
 amsmath,amssymb,
 aps,
pra,
longbibliography
]{revtex4-2}

\usepackage{hyperref}
\usepackage{notes2bib}
\usepackage{xcolor}
\usepackage{algorithm}
\usepackage{algorithmicx}

\newcommand{\e}{\mathrm{e}}
\newcommand{\ii}{\mathrm{i}}
\newcommand{\abs}[1]{\left\vert#1\right\vert}
\usepackage{braket}
\usepackage{physics}
\usepackage[a4paper, total={6.5in, 9in}]{geometry}

\usepackage[nice]{nicefrac}

\usepackage[normalem]{ulem}

\usepackage{color}

\usepackage{amsthm}
\usepackage{amsmath}
\usepackage{mathtools}

\usepackage{algpseudocode}
\usepackage{ulem}
\newcommand{\um}[1]{\,\mathrm{#1}}

\DeclareMathOperator{\sinc}{sinc}

\begin{document}

\title{High-Precision Measurement of Time Delay with Frequency-Resolved Hong-Ou-Mandel Interference of Weak Coherent States}

\author{Francesco Di Lena}
\affiliation{Agenzia Spaziale Italiana - Matera Space Center, Contrada Terlecchia snc, 75100 Matera, Italy}

\author{Fabrizio Sgobba}
\affiliation{Agenzia Spaziale Italiana - Matera Space Center, Contrada Terlecchia snc, 75100 Matera, Italy}
\affiliation{Dipartimento Interateneo di Fisica, Universit\`a di Bari, 70126 Bari, Italy}
\affiliation{INFN, Sezione di Bari, 70126 Bari, Italy}

\author{Danilo Triggiani}
\affiliation{INFN, Sezione di Bari, 70126 Bari, Italy}
\affiliation{Dipartimento Interateneo di Fisica, Politecnico di Bari, 70126 Bari, Italy}

\author{Andrea Andrisani}
\affiliation{Agenzia Spaziale Italiana - Matera Space Center, Contrada Terlecchia snc, 75100 Matera, Italy}

\author{Cosmo Lupo}
\affiliation{Dipartimento Interateneo di Fisica, Universit\`a di Bari, 70126 Bari, Italy}
\affiliation{INFN, Sezione di Bari, 70126 Bari, Italy}
\affiliation{Dipartimento Interateneo di Fisica, Politecnico di Bari, 70126 Bari, Italy}

\author{Piergiorgio Daniele}
\affiliation{Dipartimento di Elettronica, Informazione e Bioingegneria, Politecnico di Milano, 
Piazza Leonardo da Vinci 32, 20133 Milano, Italy}

\author{Gennaro Fratta}
\affiliation{Dipartimento di Elettronica, Informazione e Bioingegneria, Politecnico di Milano, 
Piazza Leonardo da Vinci 32, 20133 Milano, Italy}

\author{Giulia Acconcia}
\affiliation{Dipartimento di Elettronica, Informazione e Bioingegneria, Politecnico di Milano, 
Piazza Leonardo da Vinci 32, 20133 Milano, Italy}

\author{Ivan Rech}
\affiliation{Dipartimento di Elettronica, Informazione e Bioingegneria, Politecnico di Milano, 
Piazza Leonardo da Vinci 32, 20133 Milano, Italy}

\author{Luigi Santamaria Amato}
\email{luigi.santamaria@asi.it}
\affiliation{Agenzia Spaziale Italiana - Matera Space Center, Contrada Terlecchia snc, 75100 Matera, Italy}

\begin{abstract}
We demonstrate a scheme for high-precision measurements of time delay based on frequency-resolved Hong-Ou-Mandel (HOM) interference.
Our approach is applied to weak coherent states and exploits
an array of single-photon avalanche diodes (SPADs).
Unlike conventional HOM experiments, our setup enables high-precision measurements producing an uncertainty per coincidence of about $\sim 10$ ps even for photons separated by delays up to $\sim 4$ ps so much greater than their coherence time where ordinary non-resolved HOM fails. This result confirms our newly developed theoretical predictions that consider, differently from previous theoretical results, a finite frequency resolution in the detection.
We compare the performance of this scheme against the conventional non-resolved case. 
Experimental data align well with the predictions of quantum estimation theory, demonstrating
a significant reduction in the uncertainty.
Due to the physics of the frequency-resolved HOM effect, the gain in precision is particularly high when the estimated time delay is much longer than the coherence time.
\end{abstract}

\maketitle

\onecolumngrid
\section{Introduction}
\label{sec:intro}

Hong-Ou-Mandel (HOM) interference~\cite{Hong1987} is an intriguing quantum optical phenomenon that has become a crucial tool in many applications in quantum metrology, telecommunications, and in the wider field of quantum technology~\cite{Bouchard2020}.
In its best-known form, it describes the behavior of two indistinguishable photons that, coupled into two separate inputs of a 50:50 beam splitter (BS), bunch together and escape from the same output port.
In fact, the measurement of the average coincidences between photons exiting from different BS ports as a function of a distinguishability parameter $\eta$ shows a drop in coincidence events registered  with $\eta$ approaching 0. For example, if the only distinguishability parameter is time, the graph of coincidences versus differences in times of arrival $\Delta t$ shows a dip at $\Delta t=0$.
The dependence of the coincidence counts 
on a suitable physical parameter
(e.g.~the delay $\Delta t$) can be exploited to  measure the values difference itself (e.g.~$\Delta t$) that, in turn, is the cause of the distinguishability. 

Advances in single photon detection~\cite{DelloRusso2022,Nomerotski2020} and light engineering~\cite{Walborn2003,DAmbrosio2019,Kim2017} have strongly improved the performances of HOM-based experiments.
A pioneering work~\cite{Lyons2018} has recently reported  
attosecond precision (4.7 as) and accuracy (0.5 as).
This work, as well as other metrological applications of the HOM effect, e.g.~\cite{Harnchaiwat2020,Sgobba2024}, have exploited an optimization of the setup 
informed by quantum estimation theory.
In the ``ordinary'' HOM experiments described above, the coincidence rate approaches a constant if the dimensionless distinguishability parameter is greater than one. For example, in the case of time delay, the relevant quantity is the ratio between the delay $\Delta t$ and the biphoton coherence time $\tau$.
Consequently, traditional HOM-based interferometry is blind to quantum interference when the two photons are distinguishable at the detectors, i.e.~$\Delta t/\tau>1$.
To overcome this limitation, and due to the quantum nature of HOM interference~\cite{Pittman1996,Kwiat1992},
an initial distinguishability of the two photon pathways can be erased right before detection, fully retrieving the interference pattern (a setup historically known as "quantum eraser" \cite{Wheeler1978,Scully1982,Kim2000}).
In particular, given the parameter of interest (time of arrival in our case), resolving in the conjugate variable (frequency) restores the indistinguishability between the two photons also for values of the delay $\Delta t\gg\tau$ as a direct consequence of Heisenberg uncertainty principle.
Considerable improvements in single photon detectors and time tagging electronics capabilities have indeed made it possible to devise experimental configurations where the measurements are taken by resolving the photons' degrees of freedom. 
The potential of resolved coincidence measurement has been shown as a proof of principle ~\cite{Legero2004,Jin2015,Gerrits2015,https://doi.org/10.48550/arxiv.2104.01062}, and its reach has been extended to quantum communication~\cite{PietxCasas2023}, quantum optical coherence tomography ~\cite{IbarraBorja2019,YepizGraciano2020}, imaging~\cite{Parniak2018, Muratore2025}, entanglement generation~\cite{Zhao2014,Jin2016}, boson sampling~\cite{Orre2019,Tamma2015,Wang2018}, generation and detection of structured coalescence states~\cite{schiano2024} and precise measurement~\cite{Parniak2018,Chen2023,Triggiani2025,Maggio2025}.

For the goal of precision time measurements, an approach employing frequency entangled photons (discrete photon entanglement) through estimation of biphoton beat notes has also demonstrated promising results~\cite{Chen2019}.
However, only non-classical states of light~\cite{Chen2019,Chen2023} have been investigated in this context. 
In particular, the use of heralded photon pairs imposes strict limitations in terms of bandwidth and number of generated photon pairs.

Here we follow a different route and experimentally demonstrate a scheme recently discussed as a theoretical proposal by Triggiani et al.~\cite{Triggiani2023, Triggiani2024}.
The theory of Refs.~\cite{Triggiani2023, Triggiani2024}~shows that, for sufficiently high resolution of the detectors, resolved HOM-based measurement can be employed to estimate arbitrarily large values of the delay or the transverse displacement between the photons.
To the best of our knowledge, our work is the first demonstration of a time-delay estimation scheme with frequency-resolved HOM interference of coherent light.
In addition, and in contrast to previous works, we used, as frequency-resolving detectors, dispersive gratings and array of Single-Photon Avalanche Diodes (SPADs) on both outputs rather than filters or dispersive fibers.
The use of coherent light implies better performance in terms of photons bandwidth and generated photon rate if compared with heralded photons for example generated by pumping spontaneous parametric downconversion  crystals. Moreover the use of SPAD arrays offer better performance in term of acquisition time, since it is possible to manage a larger photon flux.
Our theoretical and experimental results show that, despite finite resolution, the frequency-resolving technique allows us to estimate time delays much larger than the coherence time of the photons without loss of sensitivity.
This result demonstrates the disruptive potential of frequency-resolved HOM metrology.

\section{Material and Methods}
\subsection{Theory}
\label{sec:theory}

We consider a two-photon interference setup equipped with single-photon frequency-resolving detectors, where two identical photons impinge on the two faces of a balanced beam splitter at a given temporal delay $\Delta t$.
The probability of detecting the frequencies $\omega_{1},\omega_{2}$ for photons bunching (B) or antibunching (A) at the output ports of the beam splitter is given by~\cite{Triggiani2023}
\begin{equation}
	P_{A/B}(\omega_{1},\omega_{2})= \frac{1}{2}f(\omega_{1})f(\omega_{2})\left(1\mp\mathcal{V}\cos((\omega_{1}-\omega_{2})\Delta t)\right),
    \label{eq:InfRes}
\end{equation}
where $f$ is the spectral distribution of the single photon, $\mathcal{V}$ is the visibility of the interference fringes, bounded to $\mathcal{V}\leqslant 0.5$ for independent coherent states\cite{Chen2016}, and $\Delta t$ is the delay between the photons.
For simplicity, we will consider $f$ as a Gaussian spectral distribution with variance $\sigma^2=\frac{1}{4\tau^2}$, with $\tau$ coherence time of the photon.
Noticeably, Eq.~\eqref{eq:InfRes} corresponds to the scenario in which the frequency resolution of the detectors, primarily determined by the pixel size of the cameras and by the resolution of the grating, and quantified by the smallest difference $\delta_\omega$ in the frequencies that can be distinguished, is large enough so that $\delta_\omega \Delta t\ll 1$ and $\delta_\omega \tau\ll 1$.
If this condition is not verified, the continuous probability density distribution in Eq.~\eqref{eq:InfRes} should be integrated over the frequency range associated to each pixel of the cameras, resulting in a discrete probability mass function
\begin{equation}
	P_{A/B}^{\omega_{01},\omega_{02}}=\int_{\omega_{01}-\frac{\delta_\omega}{2}}^{\omega_{01}+\frac{\delta_\omega}{2}}\int_{\omega_{02}-\frac{\delta_\omega}{2}}^{\omega_{02}+\frac{\delta_\omega}{2}}\dd\omega_1\dd\omega_2\ P_{A/B}(\omega_{1},\omega_{2}),
    \label{eq:Discrete}
\end{equation}
where $\omega_{0i}$, $i=1,2$, are the central frequencies associated with two given pixels.
Eq.~\eqref{eq:Discrete} exactly reproduces the more realistic scenario of finite resolution and can be computed analytically, but takes, in general, a complicated form.
We can simplify it by assuming that, within the integration intervals of width $\delta_\omega$, $f(\omega_i)$ is essentially constant, which is equivalent to requiring that $\delta_\omega/\sigma\ll 1$.
We can thus approximate 
\begin{equation}
	P_{A/B}^{\omega_{01},\omega_{02}}\simeq \frac{1}{2}f(\omega_{01})f(\omega_{02})\int_{\omega_{01}-\frac{\delta_\omega}{2}}^{\omega_{01}+\frac{\delta_\omega}{2}}\int_{\omega_{02}-\frac{\delta_\omega}{2}}^{\omega_{02}+\frac{\delta_\omega}{2}}\dd\omega_1\dd\omega_2\ \left(1\mp\mathcal{V}\cos((\omega_{1}-\omega_{2})\Delta t)\right),
\end{equation}
that, after elementary steps, reduces to
\begin{equation}
	P_{A/B}^{\omega_{01},\omega_{02}}\simeq \frac{1}{2}f(\omega_{01})f(\omega_{02})\delta_\omega^2\left(1\mp\mathcal{V}\sinc^2\left(\frac{\Delta t \delta_\omega}{2}\right)\cos((\omega_{01}-\omega_{02})\Delta t)\right),
    \label{eq:ResProb}
\end{equation}
where $\sinc(x)=\sin(x)/x$.
In the next section, we show that this approximation agrees with the experimental data collected.
On the other hand, by integrating $P_{B/A}(\omega_{1},\omega_{2})$ over the whole range of frequencies $\omega_1,\omega_2$, we can evaluate the usual probabilities of observing bunching or antibunching photons in a standard HOM interference experiment with bucket detectors, yielding the well-known, non-resolved (NR), HOM dip
\begin{equation}
	P_{A/B}^\mathrm{NR}=\frac{1}{2}(1\mp\mathcal{V}|\mathcal{F}(\Delta t)|^2),
    \label{eq:PNR}
\end{equation}
where $\mathcal{F}$ is the Fourier transform of the spectral distribution $f$.

Before proceeding with the analysis of the precision of our estimation setup, it is interesting to understand the predicted effect of a finite frequency resolution $\delta_{\omega}$ on the interference pattern at a more fundamental level.
For the delay values $\Delta t$ so that $\delta_\omega \Delta t\ll1$, Eq.~\eqref{eq:ResProb} reduces to Eq.~\eqref{eq:InfRes}.
In this scenario, the bunching and antibunching probability distributions present a constant quantum beat, ultimately modulated by the spectral distribution of the photons.
Remarkably, this is true even for $\Delta t/\tau\gg 1$, that is, for values of the delay $\Delta t$ for which a non-resolved approach would see no interference: despite the finite frequency resolution, the measurement completely erases the temporal distinguishability between the photons at the detection.
For higher values of $\Delta t$, the amplitude of the beats is reduced and the effective visibility $\mathcal{V}\sinc^2(\Delta t \delta_\omega/2)\simeq \mathcal{V}(1-\Delta t^2 \delta_\omega^2/12)$ slowly decreases with $(\Delta t \delta_\omega)^2/12$, as the finite frequency resolution still partially erases the temporal distinguishability.
Only for $\Delta t\delta_\omega \gg 1$ the beats vanish, with the probability of bunching and antibunching reducing to Eq.~\eqref{eq:PNR}.

We now analyze the theoretical bounds on the precision for the estimation of the delay $\Delta t$, first recovering some known results in recent literature.
The ultimate precision $\delta\Delta t$ achievable in the estimation of $\Delta t$ for separable photons after $N$ repeated measurements is given by the quantum Cramér-Rao bound (QCRB)~\cite{Helstrom1969, Liu2020}
\begin{equation}
	\delta\Delta t_{Q}=\frac{1}{\sqrt{N\cdot H}}=\sqrt{\frac{2}{N}}\tau,
    \label{eq:QFI}
\end{equation}
where $H$, the quantum Fisher information for separable two photon state, does not depend on the time delay to be estimated.
In appendix \ref{sec:qufi} we show that the QFI associated with our experiment coincides with $H$ in Eq.~\eqref{eq:QFI}.

Remarkably, it is known that frequency-resolved two-photon interference detection can saturate such an ultimate precision for perfect interference visibility ($\mathcal{V}=1$) and high frequency resolution ($\delta_\omega\ll\sigma$ and $\delta_\omega\Delta t\ll1$)~\cite{Triggiani2023}. 
In this case, the best precision achievable, given by the Cramér-Rao bound (CRB) $\delta\Delta t_R=\frac{1}{\sqrt{N\cdot F_\mathrm{R}}}$, where $F_{\mathrm{R}}$ is the Fisher information associated with the resolving approach~\cite{Cramer1999}, reduces to Eq.~\eqref{eq:QFI} independently of the value of the delay $\Delta t$.
For imperfect visibilities $\mathcal{V}<1$, although the optimality is lost, it is still possible to estimate delay arbitrarily larger than the coherence time of the photons with the frequency-resolving scheme~\cite{Triggiani2023}.
On the other hand, if a non-resolved measurement scheme is adopted, the best uncertainty attainable, given by the CRB $\delta\Delta t_\text{NR}=\frac{1}{\sqrt{N\cdot F_\text{NR}}}$ where $F_\text{NR}$ is the Fisher information associated with non-resolved two-photon interference measurements, can be calculated for Gaussian wave packets as~\cite{Triggiani2023}
\begin{equation}
    F_\text{NR}=\frac{H}{2}\frac{\mathcal{V}^2}{\exp{\left(\frac{\Delta t^2}{2\tau^2}\right)-\mathcal{V}^2}}\frac{\Delta t^2}{\tau^2}.
\label{eq:fnr}
\end{equation}
Noticeably, even for perfect visibility $\mathcal{V}=1$, the Fisher information $F_\mathrm{NR}$ depends on the value of the delay to be estimated, decreasing exponentially with $|\Delta t/\tau|$.
This is a well-known drawback of sensing schemes based on standard two-photon interference, since these techniques are blind to the interference between photons whose wave packets do not overlap in time.
This drawback is overcome with the use of frequency-resolving cameras, but the effect of a finite resolution remained so far an open question.

\begin{figure}
    \centering
    \includegraphics[width=0.6\linewidth]{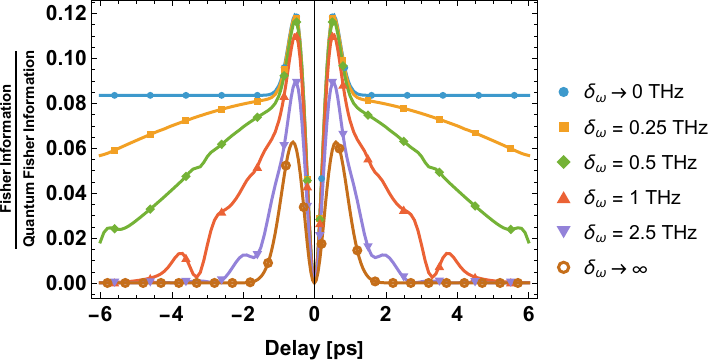}
    \caption{Plots of the Fisher information ($s^{-2}$),  normalized over the quantum Fisher information $H=2\sigma^2$ ($s^{-2}$), associated with the frequency-resolving measurement scheme for different values of the frequency resolution $\delta_\omega$, and the two limit cases of infinite resolution ($\delta_\omega\rightarrow0$) and bucket detectors ($\delta_\omega\rightarrow\infty$). We can see that the finite resolution of the frequencies at the output of the beam splitter allows the estimation of delays $\Delta t$ much larger than the coherence time $\tau$ of the photons, and it offers an overall better sensitivity than the non-resolved approach. $\mathcal{V}=0.4,\tau=0.44~\mathrm{ps}$.}
    \label{fig:FIs}
\end{figure}

We show the effect of a finite frequency resolution $\delta_\omega$ on the precision of the estimation in \figurename~\ref{fig:FIs}, where we compare the QFI $H=\frac{1}{2\tau^2}$ and the FI $F_\mathrm{R}$ and $F_\mathrm{NR}$, with the FI $F_{\delta_\omega}$ associated with a setup with finite frequency resolution $\delta_\omega$, evaluated in Appendix~\ref{app:derivation}.
We can see from \figurename~\ref{fig:FIs} that, despite a more realistic finite frequency resolution, the estimation uncertainty is smaller than the non-resolving setup for every value of the delay.
This is especially true for $\Delta t\geqslant\tau$, where $F_{\delta_\omega}$ decreases with $\Delta t$ at a much slower rate than $F_{\mathrm{NR}}$, effectively allowing for the estimation of delays that are much larger in magnitude of the coherence time of the photons.
This is our main theoretical result. We refer to Appendix~\ref{app:derivation} for the derivation of the finite frequency-resolution CRB.
In the following, we present the experimental setup that we have developed and employed to test our theoretical predictions.

\subsection{Experimental Setup}
\label{sec:exp}

\begin{figure}
\centering
\includegraphics[width=.65\columnwidth]{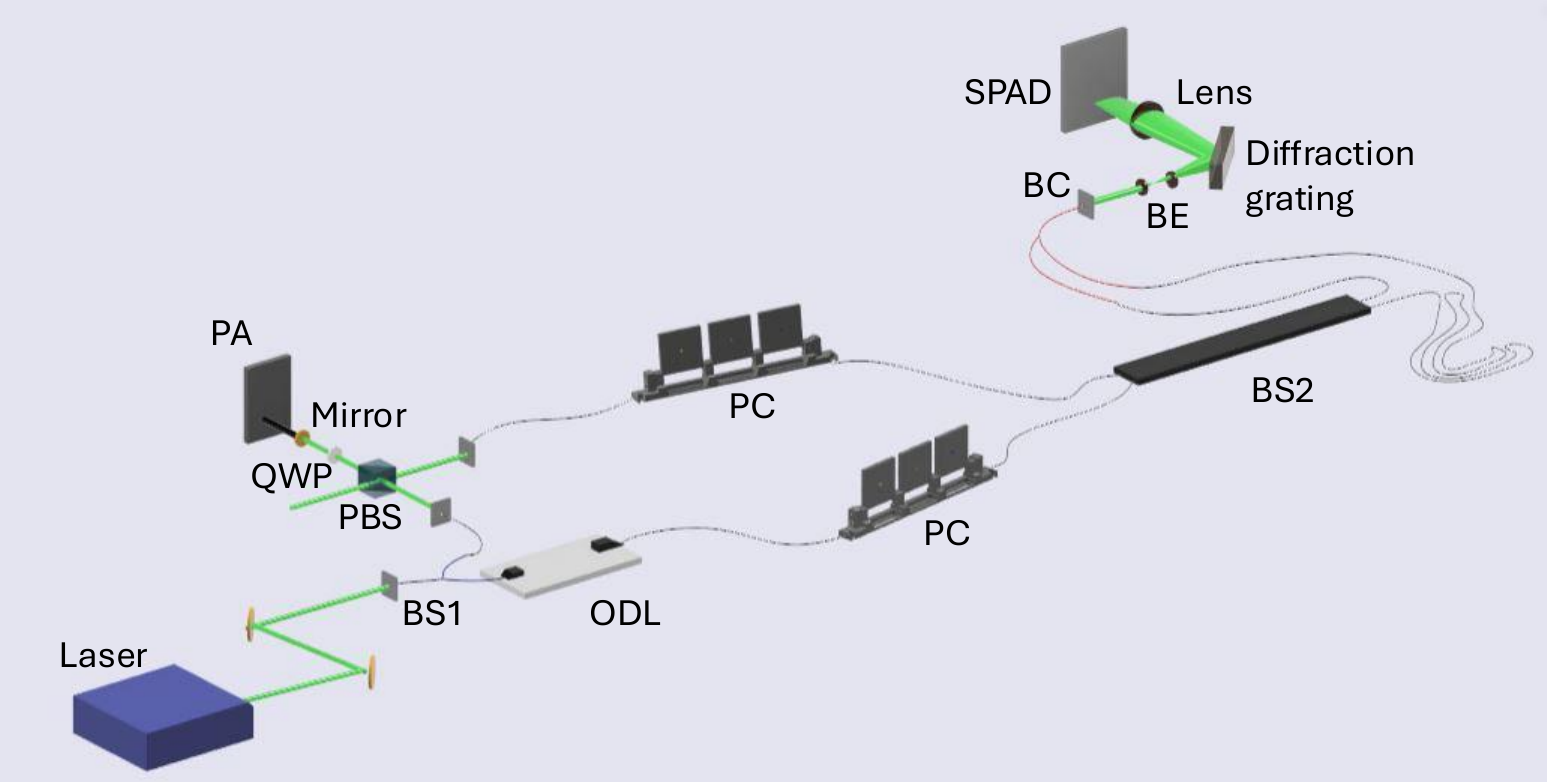}
\caption{Experimental setup of frequency resolved HOM. The light output from the laser head is split into the two arms of the MZI by the first beam splitter (BS1). In the first arm of the interferometer we have an optical delay line (ODL) followed by a polarization controller (PC), in the second one there is a polarization-based phase randomization system (PBS, quarter waveplate (QWP), mirror mounted on piezo actuator (PA)) and than a second PC. After the second beam splitter (BS2) the photons then travel through fibers of different lengths until they reach the beam combiner (BC) at a time interval of half a laser repetition time. Finally, the photons reach the diffraction grating and the SPAD array.}
\label{fig:setup}
\end{figure}

The presented optical setup relies on a Mach-Zehnder interferometer (MZI), which employs a pulsed laser as its light source. A delay line is inserted in one arm of the MZI to introduce a time-imbalance between the two arms, whereas in the other arm a moving mirror removes first order interference by randomizing the relative phase between photons through changes in the optical path. The photons exiting the two ports of the second BS travel along optical paths whose difference is set to half the repetition time of the source, to be later recombined and sent to a frequency-resolving single photon detector array, as shown in \figurename~\ref{fig:setup}.

The photon source is a pulsed laser with a central wavelength of $531.5\um{nm}$, a line width of $\sim 0.5\um{nm}$ and a repetition rate of $40\um{MHz}$. After an attenuator to ensure the photon occupation number per pulse lower than 0.1 to make multiphoton components negligible  and a half-wave plate, the laser is coupled to the single-mode fiber, part of a 50:50 fiber beam splitter (BS1).
An output of this BS passes through an optical delay line (ODL), in which photons undergo a variable path in free space and are later coupled back in fiber.
The other output is transmitted by a free-space polarizing beam splitter (PBS) and a quarter-wave plate (QWP) to be then reflected back by a mirror mounted on a piezoelectric actuator (PA) to randomize the relative phase between the photons. The reflected photons pass back through the QWP and are reflected by the PBS to be finally coupled back into an optical fiber. Prior to the second optical fiber beam splitter (BS2), the polarization controllers (PC) erase the residual distinguishability in polarization, thus maximizing the interference visibility.

The randomization of photons relative phase is obtained by driving the piezoelectric actuator with a $1\um{kHz}$ triangular wave that has peak to peak voltage $V_\text{pp}=30\um{V}$. This voltage corresponds to a change in the optical path of approximately $32\cdot\lambda/2$, and ensures, together with the choice of waveform, a uniform distribution of phases. Furthermore, with these settings, the measurement duration $t_\text{meas}\approx6.1\um{s}$ is much longer than the $31.3\um{\mu s}$ it takes the piezoelectric actuator to vary the optical path by one $\lambda/2$.

The photons exiting the BS2 travel on two optical fibers of different lengths, so that they reach a beam combiner (BC) at a time interval of about $12.5\um{ns}$, corresponding to half the repetition time of the laser. In this way, both output of BS2 go to the same identical spectrometer. We indicate the two BS outputs with ``early'' and ``late''.

The spectrometer consists of a beam expander (BE) with a magnification factor of $M=5.33$, followed by a ruled reflective diffraction grating with 1200 grooves/mm and a plano-convex lens with a focal length of $750\um{mm}$.
At the focal plane of the lens, 8 single-photon avalanche diodes (SPADs) from a 32-SPAD array, developed at the Polytechnic University of Milan~\cite{Cuccato2013}, are positioned for detection. The SPAD array features a center-to-center distance of $250\um{\mu m}$ between adjacent devices, corresponding to an expected beat frequency of $\Delta\omega\simeq1.8\um{THz}$.
The main source of uncertainty arises from the $50\um{\mu m}$ diameter of the sensitive area of the SPADs.
The SPADs exhibit a timing jitter of $45\um{ps}$ (FWHM). Their output signals are processed by a custom timing module, also developed by the same research group, operating in time-tagging mode with a resolution of $1.5\um{ps}$. Synchronization is provided by a $40\um{MHz}$ reference clock signal from the laser head.

The experiment is performed on an ordinary optical table in a thermal controlled laboratory (about 1 $^\circ C$ stability) but, due to the high acquisition speed, the stability will not pose a problem in a
real-word scenario. In fact, owing to the performance of the adopted SPAD
array, the measurement takes only a few seconds, during which the temperature
variation is completely negligible.

\section{Results and discussion}
\label{sec:meas}

The goal of our experiment is to test our theoretical prediction of the error $\delta\Delta t_{\delta_\omega}$ in the estimation of the delay 
$\Delta t$ with the finite-resolution frequency-resolving approach.
To do so, we have estimated the experimental uncertainty $\delta\Delta t_\text{exp}$ multiplied by the square root of the total number of coincidence events obtained in our setup $\sqrt{N}$, and compared it to our theoretical prediction $\delta\Delta t_{\delta_\omega}$ multiplied by $\sqrt{N}$ (whose corresponding FI is shown in  \figurename~\ref{fig:FIs}).

\begin{figure}
\centering
\includegraphics[width=1\linewidth]{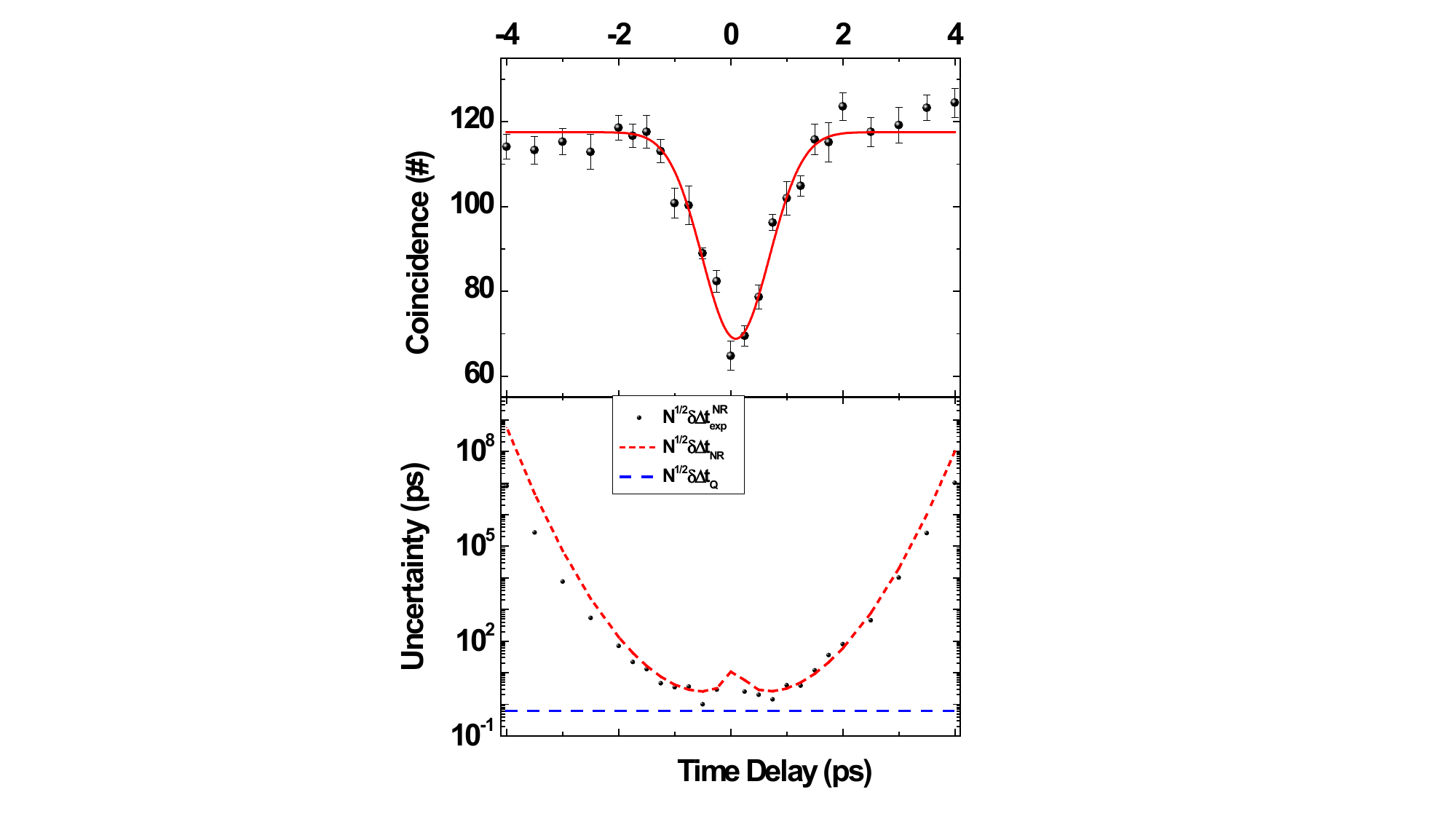}
\caption{(Upper panel) Experimental number of coincidence $C(\Delta t)$ as function of time delay between photons impinging on BS (black points) and fitted curve $C_{fit}(\Delta t)$ (red line).\\(Lower panel)  $\delta \Delta t_\text{exp}^\mathrm{NR}$ $\sqrt{N}$ (black points), theoretical uncertainty in non resolved case $\delta \Delta t_\text{NR}$$\sqrt{N}$ (red line), theoretical ultimate quantum uncertainty $\delta\Delta t_{Q}$$\sqrt{N}$ (blue line). The data represent the avarage of coincidences measured over $n_r=10$ 
repeated measurements.}
\label{NR CRB}
\end{figure}

We have at first performed a control experiment with bucket detectors, that is, without resolving the frequencies of the photon, to retrieve the already known result of the sensitivity for delay estimation in a standard HOM experiment.
In the non-resolved case, after removing the grating, we have acquired the HOM dip shown in the upper panel in \figurename~\ref{NR CRB} by recording the average coincidence count $C(\Delta t)$ on the output ports of BS2 as a function of the relative time delay $\Delta t$ between the photons impinging on BS2, controlled through an optical delay line (ODL). 
Each value of $C(\Delta t)$, at a given delay $\Delta t$, is calculated by the average of $n_r=10$  repeated measurements $c_i(\Delta t)$, while we have estimated the experimental uncertainty $\delta C$ associated with the count $C(\Delta t)$ as {$\frac{\delta c}{\sqrt{n_r}}$, where $\delta c$ is} the mean squared error of $c_i(\Delta t)$.
The upper panel of \figurename~\ref{NR CRB} shows the acquired average coincidence count $C(\Delta t)$ in a $2\um{ns}$ time window (black points) and a superimposed fitted curve $C_\mathrm{fit}\left(\Delta t\right) = A \left( 1-\mathcal{V} \exp\left(-\frac{\Delta t^2}{4\tau^2}\right) \right)$, justified by Eq.~\eqref{eq:PNR},
where the amplitude $A$, the visibility $\mathcal{V}$ and the coherence time $\tau$ are the fitting parameters.
Since our purpose is to test the theoretical prediction, which assumes perfect knowledge of all the parameters of the model, we consider the fitting parameters exact and without error.
In a real estimation procedure, this would be achieved with a one-time precise calibration of the setup prior to the estimations.

To calculate the experimental uncertainty $\delta \Delta t_{\mathrm{exp}}^\mathrm{NR}(\Delta t)$ over $N=n_pn_r$ pairs of observed photons, where $n_p=C(\Delta t)$ is the number of observed photons pairs and $n_r$ the number of repeated measurements  as a function of the delay, similarly to the procedure used in \cite{Santamaria22}, we employ the fitted curve $C_\text{fit}(\Delta t)$ as a model to estimate the delay set by the ODL by inversion.
We obtain the experimental uncertainty $\delta \Delta t_\text{exp}^{NR}$ on the measure of delay $\Delta t_\text{exp}^\mathrm{NR}$ with our non-resolving setup by error propagation from the uncertainty in the photon coincidence $C_\text{fit}$, namely

\begin{align}
    \delta \Delta t_\text{exp}^\mathrm{NR} = \left| \frac{\partial C_\text{fit}}{\partial \Delta t} \right|^{-1} \delta C .
    \label{deltatexp}
\end{align}
The lower panel of \figurename~\ref{NR CRB} shows the experimental uncertainty $\sqrt{N}\delta \Delta t_\text{exp}^\mathrm{NR}$ (black points), the expected uncertainty $\sqrt{N}\delta \Delta t_{NR}$ (red curve) given by the Fisher information $F_\mathrm{NR}$ in Eq.~\eqref{eq:fnr}, and the ultimate quantum CRB $\sqrt{N}\delta\Delta t_{Q}$ (blue curve) shown in 
Eq.\eqref{eq:QFI}, per pair of observed photons. It is clear that the uncertainty associated with the adopted non-resolved scheme rapidly diverges for large delays $\Delta t\gg\tau$, with $\tau\simeq 0.44~\mathrm{ps}$ given by the fit of $C_\text{fit}$.

\begin{figure}
\centering
\includegraphics[width=.8\linewidth]{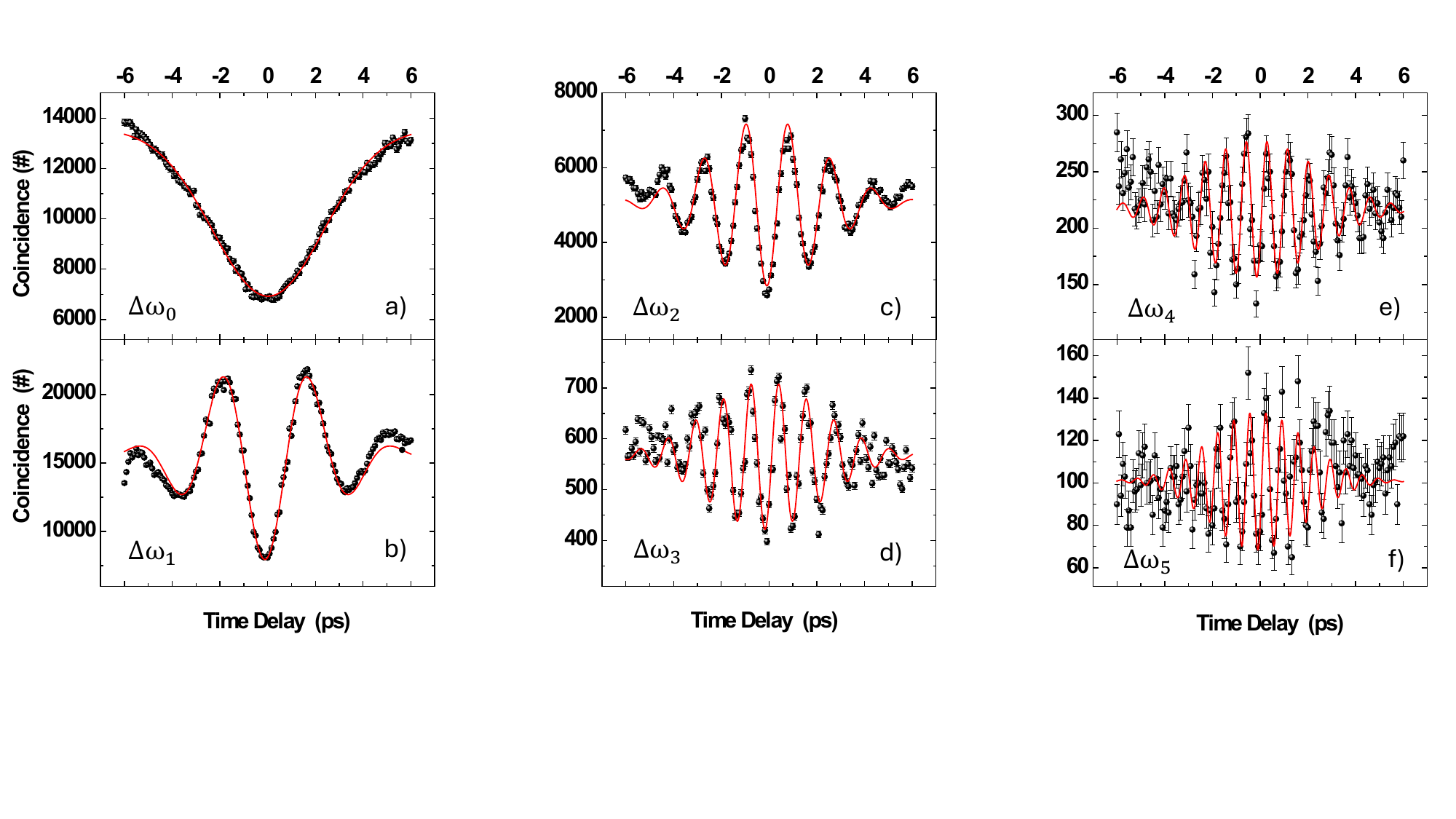}
\caption{Frequency resolved Hong-Ou-Mandel effect. In panel (a) we consider pixel 4 for both early and late port corresponding to a difference frequency equal to 0 $THz$  ($\Delta\omega_0 $). According to \cite{Triggiani2023} we observe a wider HOM dip. In the other panels we report the coincidences for different couples of early-late pixels corresponding to different detected frequencies difference. The data represent the avarage of coincidences measured over $n_r=10$ repeated measurements.}
\label{FR HOM}
\end{figure}

\begin{figure}
\centering
\includegraphics[width=.75\linewidth]{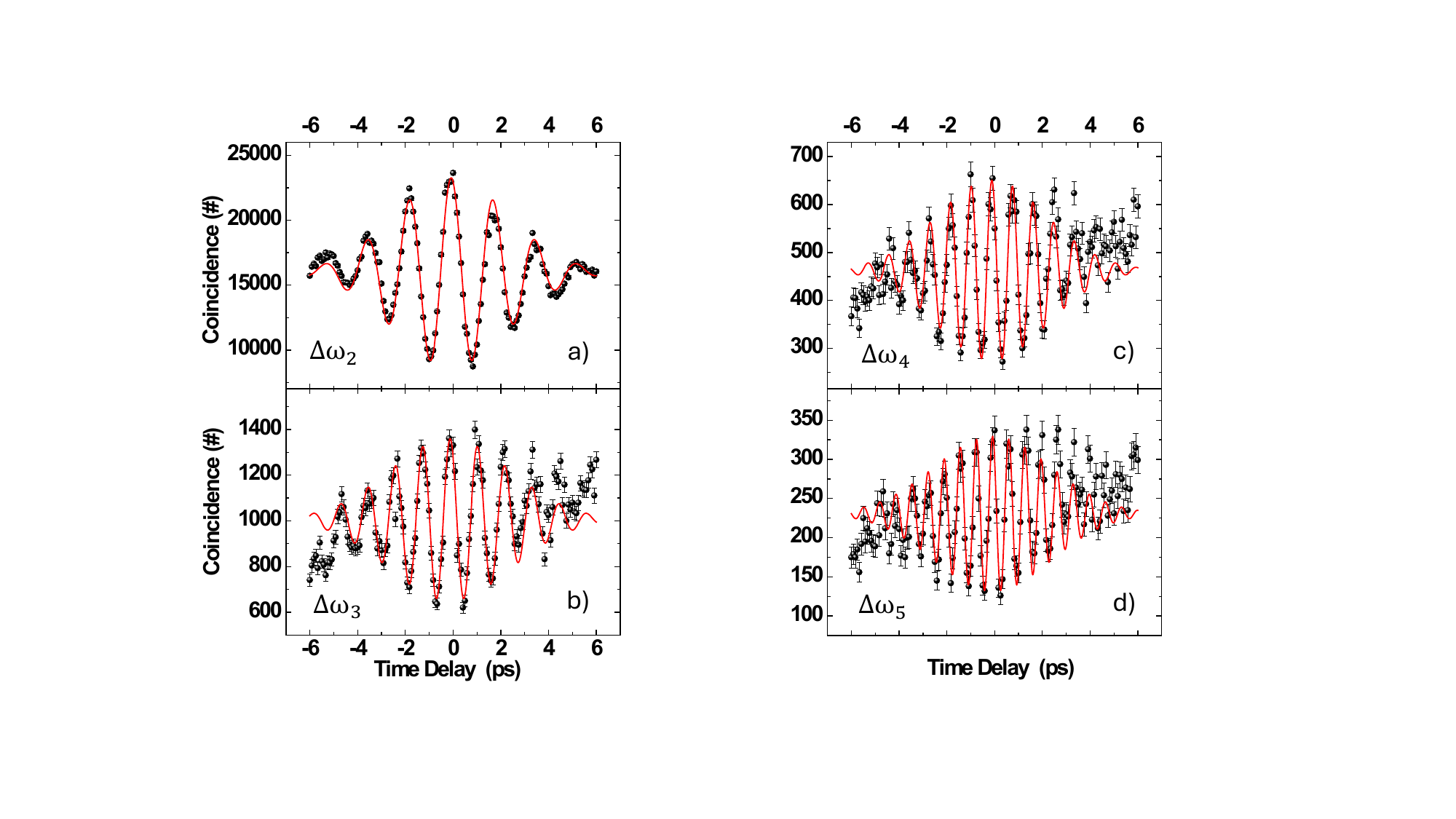}
\caption{Frequency resolved anti Hong-Ou-Mandel effect for photons coming out of the early port. In this case, correlations between adjacent SPADs are not taken into account because of cross-talk: a photon in a SPAD can create spurious avalanches also in another SPAD of the array. This effect often occurs between adjacent SPADs. The case of correlations of a SPAD with itself is not reported either. The data represent the avarage of coincidences measured over $n_r=10$ repeated measurements.}
\label{FR AHOM}
\end{figure}

With the introduction of the diffraction grating to disperse radiation, as previously explained, we have performed an acquisition using eight pixels of the SPAD array by time tagging the photons arrival in each pixels. The parameters used in the acquisition are reported in the experimental setup section. Using a specifically developed python software (see appendix~\ref{sec:code}), we calculated the matrix of coincidences $C^{B}_{i,j}$ between the pixel $i$ and $j$ around the same detectors time difference   representing the same output port of the BS2 HOM beam splitter (bunching), and matrix $C^{A}_{i,j}$ of coincidence around 12.5 ns time difference representing different output ports (anibunching). 
The coincidences are calculated using repeated measurement to estimate the experimental uncertainty as in the non resolved case. 
Each pixel $i$ of the array, as explained in the experimental setup paragraph, corresponds to a given wavelength range $\delta\lambda$ centered at $\lambda_i$ with its associated optical frequency $\omega_i$.
The black points in \figurename~\ref{FR HOM} (\ref{FR AHOM}) show the counted coincidences $C^{A}_{i,j}$($C^{B}_{i,j}$) for six(four) different values of pixel pairs $i,j$ corresponding to optical frequency pairs ($\omega_i,\omega_j$). The six (four) plots demonstrate the presence of quantum beating in the coincidence counts between pixels at different separations $k=|i-j|$, corresponding to different frequency differences $\Delta \omega_k=|\omega_i-\omega_j|$. We have fitted the data with
\begin{equation}
    C^{A/B}_{k,\text{fit}}\left(\Delta t\right) = \mathcal{N} \left( 1\mp\mathcal{V} \sinc^2\left(\frac{\Delta t \delta}{2}\right)\cos(\Delta\omega_k\Delta t)\right),
    \label{eq:ResFit}
\end{equation}
justified by our model in Eq.~\eqref{eq:ResProb}.
The fits produces beating frequency $\Delta \omega_k$, that we report in the table, proportional to $k$ as expected. 
From \figurename~\ref{FR HOM} and \figurename~\ref{FR AHOM}
it is clear that the oscillations are still visible outside the coherence time $\tau \simeq 0.44~\mathrm{ps}$, which we show enables the estimation of time delays even for photons that do not overlap in time.
Similarly to the analysis for the non-resolving setup, we consider the fitted curves in Eq.~\eqref{eq:ResFit} as exact and without error, since we aim to compare the experimental uncertainties with the associated CRB, which assumes perfect knowledge on the parameters of the model.

We determined the experimental uncertainty $\delta \Delta t_\text{exp}$ for the frequency-resolving case through a maximum-likelihood estimation approach.
Employing the fitted curves in Eq.~\eqref{eq:ResFit} as probability mass function generating the measured data $C^X_k$, with $X=A,B$ and $k=|i-j|$, we evaluate the log-likelihood function
\begin{equation}
	\mathcal{L}(\Delta t)=\sum_{\substack{X=A,B\\k}}C_k^X\ln C^X_{k,\mathrm{fit}}(\Delta t)
    \label{eq:LogLike}
\end{equation}
and thus the uncertainty over $N=n_r \sum\limits_{\substack{X=A,B\\k}}C_k^X$ observed coincidences associated with the value of $\Delta t$ that maximizes $\mathcal{L}(\Delta t)$ is given by
\begin{equation}
	\delta \Delta t_\mathrm{exp}=\sqrt{\sum\limits_{\substack{X=A,B\\k}}\left(\delta C_k^X\frac{\frac{\partial^2\mathcal{L}(\Delta t)}{\partial C_k^X\partial\Delta t}}{\frac{\partial^2\mathcal{L}}{\partial\Delta t^2}}\right)^2},
    \label{eq:UncRes}
\end{equation}
caused by the fluctuations $\delta C_k^X$ of the experimental data $C_k^X$ (more details in Appendix~\ref{app:Likelihood}).
\figurename~\ref{FR:CRB} shows $\sqrt{N}\delta \Delta t_\text{exp}$ (black points), quantum CRB (blue line), CRB for the non-resolved approach as obtained in Eq.~\eqref{eq:fnr}  (red line), and the CRB for finite frequency resolution that we developed (green line), per pair of observed photons.
The precision obtained in our frequency-resolving setup does not diverge as rapidly as the non-resolving approach confirming a good time resolution also for non-overlapping pulses as predicted by our theoretical model.

Some values of $\delta\Delta t_\mathrm{exp}$$\sqrt{N}$ lie below the theoretical bounds of the CRB and the QCRB.
This behavior can be easily explained with two observations.
First, retrieving the fitted parameters in Eq.~\eqref{eq:ResFit} and performing the estimation of the uncertainties from the same data renders our model overfitted and the estimated uncertainties too optimistic.
Second, the expression in Eq.~\eqref{eq:UncRes} has a more convoluted dependence on the expected coincidence counts compared to Eq.~\eqref{deltatexp}. In conjunction with a more delicate measurement procedure, this caused larger statistical fluctuations.
However, these complications are expected to affect equally the estimated value of $\delta\Delta t_\mathrm{exp}$ independently of the true value of the delay $\Delta t$.
In other words, the overall dependence of $\delta\Delta t_\mathrm{exp}$ on increasing values $\Delta t$, that is the focus of our study and our main result, is not affected.

\bigskip
\begin{center}
\begin{tabular}{|c|c|c|c|c|c|c|}
  \hline
   & $\Delta \omega_0\ [\mathrm{THz}]$  & $\Delta \omega_1\ [\mathrm{THz}]$ & $\Delta \omega_2\ [\mathrm{THz}]$ & $\Delta \omega_3\ [\mathrm{THz}]$ & $\Delta \omega_4\ [\mathrm{THz}]$ & $\Delta \omega_5\ [\mathrm{THz}]$  \\
    \hline
Bunching & //  & // & 3.57$\pm$0.01 & 5.49$\pm$0.03 & 7.24$\pm$0.03  & 9.30$\pm$0.03 \\
   \hline
Antibunching & set=0  & 1.72$\pm$0.01 & 3.56$\pm$0.01 & 5.42$\pm$0.03 & 7.26$\pm$0.03  & 9.22$\pm$0.06 \\
  \hline
\end{tabular}
\end{center}

\begin{figure}
\centering
\includegraphics[width=.65\linewidth]{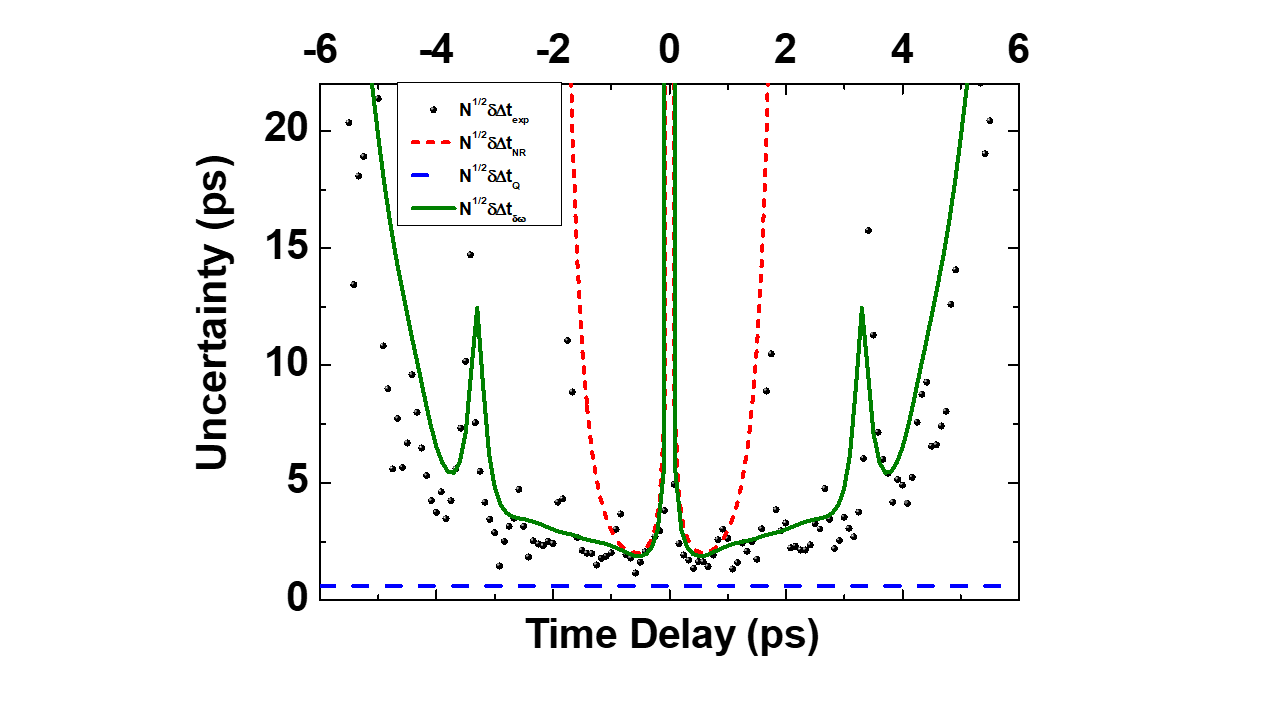}
\caption{Uncertainties comparison: Frequency resolved experimental uncertainty ($\delta \Delta t_\text{exp}$$\sqrt{N}$) calculated through maximum-likelihood estimation (black points), qCRB multiplied by $\sqrt{N}$ (blue line), NR CRB multiplied by $\sqrt{N}$ (red line), FR CRB multiplied by $\sqrt{N}$ with finite frequency resolution $\delta_{\omega} = 1$ THz (green line).}
\label{FR:CRB}
\end{figure}

\section{Conclusion}

We have proposed an experimental setup capable to perform frequency-resolved Hong-Ou-Mandel interferometry and proved its reliability with classical light. 
We have validated pioneering theoretical works~\cite{Triggiani2023,Triggiani2024} and showcased the metrological
advantage achievable with the frequency-resolved configuration in realistic experimental conditions, i.e.~with finite resolutions, while performing time-delay estimations. 

The results reported, far from being limited to the demonstration of intriguing fundamental properties of light, have the potential to become a benchmark for innovations in a variety of applications beyond metrology, 
such as boson sampling, imaging and sensing.

\appendix

\section{Derivation of the quantum Fisher information\label{sec:qufi}}
In this appendix we will evaluate the quantum Fisher information $H$ associated with the estimation of the delay $\Delta t$ found in the main text.
We consider two coherent states $\ket{\alpha}_{j}$, $j=1,2$, of the type
\begin{equation}
	\ket{\alpha}_{j}=\e^{-\frac{\abs{\alpha}^2}{2}}\sum_{n=0}^\infty \frac{\alpha^n}{n!}\hat{a}_{j,t_j}^{\dag n}\ket{0},\qquad \hat{a}_{j,t_j}^{\dag}=\int\dd \omega\ \e^{-\ii \omega t_j} \psi(\omega)\hat{a}_j^{\dag}(\omega),
    \label{eq:CohState}
\end{equation}
where $\hat{a}^\dag_{j,t_j}$ is the bosonic operator creating a photon at the $j$th input mode of the beam splitter at a time $t_j$ with respect to a certain reference time, with $t_1-t_2=\Delta t$, $\hat{a}^\dag_j(\omega)$ is the bosonic operator creating a photon with frequency $\omega$ at the $j$th input mode, and $\psi(\omega)$ is the spectral single-photon amplitude of each photon.
Let us now introduce a randomized phase $\phi_j$ that transforms the pure coherent state $\ket{\alpha}_{j}$ into the classically averaged state
\begin{multline}
	\hat{\rho}_{j}=\frac{1}{2\pi}\int_0^{2\pi}\dd \phi_j\ \ketbra{\e^{\ii\phi_j}\alpha}_{j}=\frac{1}{2\pi}\int_0^{2\pi}\dd \phi_j\ \e^{-\abs{\alpha}^2} \sum_{n,m=0}^\infty \frac{\alpha^{n}\alpha^{*m}}{n!m!}\e^{\ii\phi_j(n-m)}\hat{a}_{j,t_j}^{\dag n}\ketbra{0}{0}\hat{a}_{j,t_j}^{m}\\
    =\e^{-\abs{\alpha}^2}\sum_{n=0}^\infty \frac{\abs{\alpha}^{2n}}{n!^2}\hat{a}_{j,t_j}^{\dag n}\ketbra{0}{0}\hat{a}_{j,t_j}^{n}=\sum_{n=0}^\infty p_n \ketbra{n}{n}_{j,t_j},\quad p_n=\e^{-\abs{\alpha}^2} \frac{\abs{\alpha}^{2n}}{n!}
\end{multline} 
We notice that the dependence on $t_j$ appears only through the orthogonal states $\ket{n}_{t_j}=\frac{\hat{a}_{j,t_j}^{\dag n}}{\sqrt{n!}}\ket{0}$ which remain orthogonal after differentiation with respect to $t_j$.
We further assume that only $2$-photon events are observed through post-selection from the global state $\hat{\rho}_1\otimes\hat{\rho}_2$, since in our experiment we only collect data for such events.
After normalization, the post-selected global state becomes
\begin{equation}
	\hat{\rho}^{2ph}=\frac{1}{2}\ketbra{1}_{1,t_1}\otimes\ketbra{1}_{2,t_2}+\frac{1}{4}\ketbra{2}_{1,t_1}\otimes\ketbra{0}_{2,t_2}+\frac{1}{4}\ketbra{0}_{1,t_1}\otimes\ketbra{2}_{2,t_2}
    \label{eq:2phState}
\end{equation}
The additivity over orthogonal sectors and over tensor products of the quantum Fisher information matrix~\cite{Liu2020} allows us to write the quantum Fisher information matrix $\mathcal{H}\equiv\mathcal{H}[\hat{\rho}^{2ph}]$ associated with $t_1,t_2$ as
\begin{equation}
	\mathcal{H}=\begin{pmatrix}
	    \frac{1}{2}H[\ket{1}_{1,t_1}] +\frac{1}{4}H[\ket{2}_{1,t_1}]& 0\\
        0 & \frac{1}{2}H[\ket{1}_{2,t_2}+\frac{1}{4}H[\ket{2}_{2,t_2}]]
	\end{pmatrix}
    \label{eq:QFIM2}
\end{equation}
where $H_{t_j}[\ket{n}_{j,t_j}]$ is the quantumm Fisher information of the $n=1,2$-photon state associated with $t_j$, $j=1,2$.
The terms inside Eq.~\eqref{eq:QFIM2} can be evaluated by definition as
\begin{equation}
	H_{t_j}[\ket{n}_{j,t_j}]=4\left(\braket{\partial_j n}_{j,t_j}-\abs{\braket{n}{\partial_j n}_{j,t_j}}^2\right),\qquad \partial_j\equiv\frac{\dd}{\dd t_j}
    \label{eq:QFIDef}
\end{equation}
Employing the definition of $\hat{a}_{j,t_j}^\dag$ in Eq.~\eqref{eq:CohState} We can easily evaluate
\begin{align}
    \nonumber
    \ket{n}_{j,t_j}&=\frac{\hat{a}_{j,t_j}^{\dag n}}{{\sqrt{n!}}}\ket{0}=\frac{1}{\sqrt{n!}}\int\dd \omega_1\dots\dd\omega_n\ \e^{-\ii(\omega_1+\dots+\omega_n)t_j}\psi(\omega_1)\dots\psi(\omega_n)\times\\
    &\times\hat{a}^\dag_{j}(\omega_1)\dots\hat{a}^\dag_{j}(\omega_n)\ket{0}\\
    \nonumber
	\ket{\partial_j n}_{j,t_j}&=\frac{1}{\sqrt{n!}}\int\dd \omega_1\dots\dd\omega_n\ -\ii(\omega_1+\dots+\omega_n)\e^{-\ii(\omega_1+\dots+\omega_n)t_j}\psi(\omega_1)\dots\psi(\omega_n)\times\\
    &\times\hat{a}^\dag_{j}(\omega_1)\dots\hat{a}^\dag_{j}(\omega_n)\ket{0}
\end{align}
and thus, employing the commutation relation $[\hat{a}_j(\omega),\hat{a}^\dag_j(\omega')]=\delta(\omega-\omega')$ and the definition in Eq.~\eqref{eq:QFIDef}, we get, for generic $n$
\begin{align}
	\braket{\partial_j n}_{j,t_j}&=\int\dd \omega_1\dots\dd\omega_n\ (\omega_1+\dots+\omega_n)^2\abs{\psi(\omega_1)\dots\psi(\omega_n)}^2=n\langle\omega^2\rangle+n(n-1)\langle\omega\rangle^2\\
    \abs{\braket{n}{\partial_j n}_{j,t_j}}^2&=\abs{\int\dd \omega_1\dots\dd\omega_n\ (\omega_1+\dots+\omega_n)\abs{\psi(\omega_1)\dots\psi(\omega_n)}^2}^2=n^2\langle\omega\rangle^2\\
    H_{t_j}[\ket{n}_{j,t_j}]&=4n(\langle\omega^2\rangle-\langle\omega\rangle^2)=4n\sigma^2,
    \label{eq:QFIN2}
\end{align}
where $\langle\cdot\rangle$ denotes the expected value over the single-photon spectral distribution $\abs{\psi(\omega)}^2$.
We can use the expression of $H_{t_j}[\ket{n}_{j,t_j}]$ in Eq.~\eqref{eq:QFIN2} for $n=1,2$ to evaluate the terms in the quantum Fisher information matrix in Eq.~\eqref{eq:QFIM2}, which yields 
\begin{equation}
	\mathcal{H}=4\sigma^2I_2,\qquad \mathcal{H}^{-1}=\tau^2I_2
\end{equation}
with $I_2$ the $2\times2$ identity matrix, and $\tau=1/2\sigma$ coherence time of each photon.
In order to evaluate the quantum Fisher information associated with $\Delta t$, it is necessary to perform a change of parametrization $\Delta t=\Delta t(t_1,t_2)=t_1-t_2$.
The Jacobian vector $J=(\partial_1 \Delta t, \partial_2 \Delta t)=(1,-1)$ allows us to retrieve, from the optimal covariance matrix $\mathcal{H}^{-1}$ for $t_1$ and $t_2$, the quantum Cramér-Rao bound associated with $\Delta t$ after the observation of $N$ copies of the two-photon state in Eq.~\eqref{eq:2phState}, i.e. after $N$ pairs of observed photons,
\begin{equation}
	\delta\Delta t_Q =\sqrt{\frac{J\mathcal{H}^{-1}J^T}{N}}=\sqrt{\frac{2}{N}}\tau,
\end{equation}
as shown in the main text.

\section{Derivation of the frequency-resolved CRB for finite resolutions}
\label{app:derivation}

The Fisher information $F$ associated with the estimation of a parameter $\phi$ from the measurement of a random variable $X$ taking values $x$ from a probability function $P(x|\phi)$ is given by~\cite{Cramer1999}
\begin{equation}
	F = \mathbb{E}_X\left[\left(\frac{\partial}{\partial\phi}\ln P(X|\phi)\right)^2\right],
\end{equation}
where $\mathbb{E}_X$ denote the expectation value over the random variable $X$.
Applied to our exact probability mass function $P_{A/B}^{\omega_{01},\omega_{02}}$ found in Eq.~\eqref{eq:Discrete}, we get
\begin{equation}
	F_{\delta_\omega}=\sum_{\substack{\omega_{01},\omega_{02}\\X=A,B}}\frac{1}{P_{A/B}^{\omega_{01},\omega_{02}}}\left(\frac{\partial P_{A/B}^{\omega_{01},\omega_{02}}}{\partial\Delta t}\right)^2,
    \label{eq:FIdeltaApp}
\end{equation}
where the summation over $\omega_{01},\omega_{02}$ is performed over the central frequencies of all the pixels in the camera, which are assumed to be enough in number to cover the whole spectrum $f(\omega)$ of the photons.
Although possible, the analytical evaluation of $F_{\delta_\omega}$ from Eq.~\eqref{eq:FIdeltaApp} is cumbersome and does not yield an informative expression.
The plot of the Fisher information $F_{\delta_\omega}$ in \figurename~\ref{fig:FIs} has been numerically retrieved by setting the central frequencies equally distant $\{\omega_{01},\omega_{02}\}=\{n\delta_\omega,m\delta_\omega\}$ and summing over $n,m\in\{-5,-4,\dots,4,5\}$, which covers the whole spectrum $f$ for $\delta_\omega=1$THz and $\tau=\frac{1}{2\sigma}=0.44~\mathrm{ps}$.

\section{Software for the calculation of the coincidence}\label{sec:code}

The code to detect coincidences is written in Python.
The algorithm~\ref{alg:coinc} reports the pseudocode, where $t_a$ and $t_b$ are the vectors containing the temporal tags of the photons arriving on the two pixels, or groups of pixels, between which the coincidences are calculated.
At each step of the while loop, a coincidence is detected only if $|\Delta t|<\text{cw}$, namely if the absolute value of the difference between the arrival time of the two photons is smaller than the coincidence window. If this happens, the code considers two photons ``used'' and moves on to the next ones in both lists (lines 11$-$12).
If the coincidence condition is not met, then the script only discards the photon that arrived first (lines 14 or 16). For example, if the photon that arrived at time $t_a[i]$ is the first of the two and there is no coincidence, in the next step the coincidence between $t_a[i+1]$ and $t_b[j]$ will be verified.

\begin{figure}[h!]
\centering\begin{minipage}{0.75\linewidth}
\begin{algorithmic}[1]
\State $\text{cw} \gets 2\um{ns}$  \Comment{Coincidence window}
\State $n_a \gets \text{length}(t_a)$
\State $n_b \gets \text{length}(t_b)$
\State $cc \gets 0$ \Comment{Coincidence count}
\State $i \gets 1$
\State $j \gets 1$
\While{$i < n_a$ \textbf{and} $j < n_b$} \Comment{Ends when the first of $t_a$ and $t_b$ runs out}
    \State $\Delta t \gets t_b[j] - t_a[i]$ \Comment{Defines the time difference}
    \If{$|\Delta t| < \text{cw}$}
        \State $cc \gets cc + 1$ \Comment{Coincidence found}
        \State $i \gets i + 1$ \Comment{Consider next photon of $t_a$}
        \State $j \gets j + 1$ \Comment{Consider next photon of $t_b$}
    \ElsIf{$\Delta t > 0$}
        \State $i \gets i + 1$ \Comment{Consider next photon of $t_a$}
    \Else
        \State $j \gets j + 1$ \Comment{Consider next photon of $t_b$}
    \EndIf
\EndWhile
\end{algorithmic}
\end{minipage}
\caption{Counting coincidences between two time vectors $t_a$ and $t_b$}
\label{alg:coinc}
\end{figure}

\section{Maximum-likelihood estimator and sensitivity in Eq.~\eqref{eq:UncRes}}
\label{app:Likelihood}

The maximum-likelihood estimator $\Delta t_\mathcal{L}(\{C_k^X\})$ is defined as the value of $\Delta t$ that, given the set of coincidence counts $\{C_k^X\}$ observed, maximizes the likelihood function
\begin{equation}
	\mathcal{L}(\Delta t)=\sum_{\substack{X=A,B\\k}}C_k^X\ln C^X_{k,\mathrm{fit}}(\Delta t),
    \label{eq:DerLogApp}
\end{equation}
already shown in Eq.~\eqref{eq:LogLike}, with $k$ running over the set of pixel pairs, where we assumed that $C^X_{k,\mathrm{fit}}$ are the exact probabilities that generated the coincidence counts $\{C_k^X\}$.
Then, by definition,
\begin{equation}
	0=\frac{\partial \mathcal{L}(\Delta t)}{\partial \Delta t}\Bigg\vert_{\Delta t=\Delta t_\mathcal{L}}=\sum_{\substack{X=A,B\\k}}C_k^X\frac{\partial}{\partial\Delta t}\ln C^X_{k,\mathrm{fit}}(\Delta t)\Bigg\vert_{\Delta t=\Delta t_\mathcal{L}}.
\end{equation}
If we differentiate this expression with respect to the outcomes $C_k^X$, remembering that $\Delta t_\mathcal{L}\equiv\Delta t_\mathcal{L}(\{C_k^X\})$ depends on the outcomes, we have
\begin{equation}
	0=\frac{\dd}{\dd C_k^X} \frac{\partial \mathcal{L}(\Delta t)}{\partial \Delta t}\Bigg\vert_{\Delta t=\Delta t_\mathcal{L}} =\frac{\partial^2 \mathcal{L}(\Delta t)}{\partial C_k^X\partial \Delta t}\Bigg\vert_{\Delta t=\Delta t_\mathcal{L}} +\frac{\dd \Delta t_\mathcal{L}}{\dd C_k^X}\frac{\partial^2\mathcal{L}(\Delta t)}{\partial \Delta t^2}\Bigg\vert_{\Delta t=\Delta t_\mathcal{L}}.
\end{equation}
We can thus evaluate
\begin{equation}
	\frac{\dd \Delta t_\mathcal{L}}{\dd C_k^X}=-\frac{\frac{\partial^2 \mathcal{L}(\Delta t)}{\partial C_k^X\partial \Delta t}}{\frac{\partial^2\mathcal{L}(\Delta t)}{\partial \Delta t^2}}\Bigg\vert_{\Delta t=\Delta t_\mathcal{L}},
\end{equation}
that can be easily calculated from Eq.~\eqref{eq:DerLogApp}, and thus get
\begin{equation}
	\delta \Delta t_{\mathrm{exp}} = \sqrt{\sum_{\substack{X=A,B\\k}}\left(\frac{\dd \Delta t_\mathcal{L}}{\dd C_k^X}\delta C_k^X\right)^2,}
\end{equation}
found in Eq.~\eqref{eq:UncRes}.

\vspace{0.5cm}
\noindent
\textbf{Funding.}
European Union - Next Generation EU:
NRRP Initiative, Mission 4, Component 2, Investment 1.3 - 
D.D. MUR n. 341 del 15.03.2022 – Next Generation EU (PE0000023 - "National
Quantum Science and Technology Institute).
European Union - Next Generation EU, Missione 4 Componente 1, PRIN 2022, project title "QUEXO", CUP: D53D23002850006.
Italian Space Agency (Subdiffraction
Quantum Imaging SQI 2023-13-HH.0).\\
\textbf{Acknowledgments.}
We wish to acknowledge Vincenzo Buompane, Paolo Baiocco and Graziano Spinelli for technical
support.\\

\bibliography{sample}
\end{document}